\documentclass[fleqn,twoside]{article}
\usepackage{amssymb} 

\usepackage{aas_macros,amsmath,cite,graphicx,mathtools,setspace,subfigure}
\usepackage[margin=1.3in,letterpaper]{geometry}
\usepackage[colorlinks=true,citecolor=teal]{hyperref}
\usepackage[affil-it]{authblk}

\usepackage{graphicx}
\usepackage{xcolor} 
\usepackage{bm} 
\usepackage{amsfonts}

\def\p{\partial }
\def \cO{{\cal O}}
\def \cI{{\cal I}} 
\def \cL{{\cal L}} 

\def\beq{\begin{equation}}
\def\eeq{\end{equation}}
\def\ba{\beq\begin{array}{c}}
\def\ea{\end{array}\eeq}

\def\p{\partial}

\def\d{{\delta}}

\begin{document}
\setcounter{tocdepth}{2}

\title{Celestial Holography: Lectures on Asymptotic Symmetries}

\date{}

\author[1,2]{P. B. Aneesh}
\author[3]{Geoffrey Comp\`{e}re\footnote{Corresponding author: geoffrey.compere@ulb.be}}
\author[4]{Leonardo Pipolo de Gioia}
\author[5]{Igor Mol}
\author[6]{Bianca Swidler}

\affil[1]{\small Chennai Mathematical Institute, H1 SIPCOT IT Park, Siruseri, Kelambakkam 603103, India }
\affil[2]{\small Institute of Mathematical Sciences, Homi Bhabha National Institute, IV Cross Road, C. I. T. Campus, Taramani, Chennai 600113, India}
\affil[3]{\small
Universit\'{e} Libre de Bruxelles and International Solvay Institutes\\
Brussels, CP 231 B-1050, Belgium}
\affil[4]{\small University of Campinas - UNICAMP, Institute of Physics ``Gleb-Wataghin'',
13083-859, Campinas - SP, Brazil}
\affil[5]{\small Institute for Pure and Applied Mathematics \emph{(IMPA)} \\ CEP 22460-320, Rio de Janeiro, RJ - Brasil}
\affil[6]{\small Independent Scholar, formerly Department of Physics, Princeton University, Jadwin Hall, Washington RD., Princeton, NJ 08544, USA}

\maketitle

\begin{abstract}
The aim of these Lectures is to provide a brief overview of the subject of asymptotic symmetries of gauge and gravity theories in asymptotically flat spacetimes as background material for celestial holography. 
\vspace{1pc}
\end{abstract}

\tableofcontents

\section{General considerations} \label{sec:general}

This review is intended to provide background material for the study of celestial holography for asymptotically flat spacetimes. Recent reviews on celestial holography can be found in \cite{Raclariu:2021zjz,Pasterski:2021rjz}. Complementary reviews on asymptotic symmetries include \cite{Strominger:2017zoo,Compere:2018aar,Ruzziconi:2019pzd}.

\subsection{Defining the asymptotic symmetry group} \label{subsec:Defining-Asymp-Group}

Let us start by asking what are the theories of interest where an asymptotic symmetry group can be defined. First, we need to specify the kinematics: a spacetime with either or both asymptotic boundaries and finite boundaries. In these lectures we will mainly consider Minkowski spacetimes and asymptotically flat spacetimes when gravity is present.  Second, the bulk dynamics is specified by a class of Lagrangians that are assumed to exist but which can be nonunique due to field redefinitions or dualities. Particular classes of Lagrangians are  ``gauge theories'', i.e. Lagrangians that admit non-trivial Noether identities among their equations of motion.  Because of Noether's second theorem, it is equivalent to the statement that there exist local variations of the fields that keep the Lagrangian invariant up to boundary terms. Note that because of dualities, a class of Lagrangians describing the same theory might contain gauge theories and non-gauge theories. We will be interested in formulations that are gauge theories because of the presence of local gauge transformations with associated locally constructed canonical charges (in non-gauge theory formulations they would be nonlocal). Third, the ``boundary dynamics'' or boundary conditions need to be specified. They are required at spatial and null boundaries in order to define the variational principle. In addition, consistency needs to be enforced between distinct boundaries at their intersecting ``corners''. 

There exists several frameworks to describe boundary conditions of gauge theories. The main ones are the Hamiltonian formalism \cite{Arnowitt:1962hi,Regge:1974zd,Brown:1986ed}, the Lagrangian formalism \cite{Abbott:1981ff,Lee:1990nz,Barnich:2001jy} and the Hamilton-Jacobi formalism \cite{Brown:2000dz}. There are also two distinct approaches to formulate boundary conditions: in one of them, one fixes the gauge, which needs to be by definition associated with zero canonical charges, otherwise one would miss canonical charges! This has the advantage of allowing efficient computations. This is the approach used by Bondi, van den Burg, Metzner and Sachs at null infinity for asymptotically flat spacetimes \cite{Bondi:1962px,Sachs:1962wk} (additional missed charges were found later on \cite{Barnich:2010eb,Campiglia:2014yka}). Alternatively, one only writes geometrical expressions valid in any gauge but with background structures, as done by Penrose at null infinity in asymptotically flat spacetimes \cite{Penrose:1962ij}, see also \cite{Ashtekar:1978aa,Ashtekar:1984,Ashtekar:2014zsa}. An intermediate approach is to formulate boundary conditions in specific coordinates, but which still admit gauge redundancies, e.g. the derivation \cite{Henneaux:1985ey} for asymptotically anti-de Sitter spacetimes. In all cases, the asymptotic symmetry group is defined as the quotient of the group of residual gauge transformations modulo the group of trivial gauge transformations, 
\begin{equation}\label{ASG}
    \text{Asymptotic Symmetry Group} = \frac{\text{Group of residual gauge transformations}}{\text{Group of trivial gauge transformations}}.
\end{equation}
Here trivial means that the gauge transformation is associated to a vanishing canonical charge. The asymptotic symmetry group is equivalently defined as the group of global symmetries of the class of theories given the set of boundary conditions. 

There are however some loopholes in the definitions above that we need to address. First, equivalent theories might admit distinct gauge groups. Let us take examples. Einstein gravity or the theory of the interacting spin 2 massless field admits many formulations. One of them is the metric formulation with field $g_{\mu\nu}$ and with the diffeomorphism group as gauge group. Another formulation is the Cartan formulation with fields $e^\mu_a$, $\Gamma_{b\mu}^a$ with diffeomorphisms and local Lorentz transformations as gauge transformations. In that particular case, the local Lorentz transformations are not associated to further charges \cite{Barnich:2016rwk} but that requires a computation! In the (linear) spin 1 case or electromagnetism, the standard Lagrangian formulation is the gauge potential formulation in terms of the gauge potential $A_\mu$ and associated field strength $F_{\mu\nu}=2\partial_{[\mu}A_{\nu]}$. There is however a dual gauge potential formulation obtained from expressing the Hodge dual of the field strength $G_{\mu\nu }\equiv\frac{1}{2}\epsilon_{\mu\nu\alpha\beta}F^{\alpha\beta}$, which is closed outside sources as a result of Maxwell's equations. It implies that it can be locally written as the total derivative a dual gauge field, $G_{\mu\nu}=2\partial_{[\mu}B_{\nu]}$. The gauge group acts on $A_\mu$ as $\delta A_\mu = \partial_\mu  \phi$ where $\phi=\phi(x^\mu)$ is arbitrary while the dual gauge group acts on $B_\mu$ as $\delta B_\mu = \partial_\mu \psi$ where $\psi(x^\mu)$ is arbitrary. As it turns out, the constant gauge parameter $\phi(x^\mu)=\phi$ is canonically associated with the electric monopole charge while the constant gauge parameter $\psi(x^\mu)=\psi$ is canonically associated with the magnetic monopole charge. For the spin 0 field, the standard Lagrangian formulation is in terms of a scalar field which admits a global shift symmetry of the kinetic term, but the theory admits a dual formulation in terms of an antisymmetric 2-form. The Lagrangian admits the equivalent forms \cite{Campiglia:2018see}
\begin{equation}
L=\partial^\mu \phi \partial_\mu \phi = Z_\mu Z^\mu + B_{\alpha\beta}\epsilon^{\alpha\beta \mu \nu}\partial_\mu Z_\nu = \frac{1}{4}\partial_{[\alpha}B_{\mu\nu]}\partial^\alpha B^{\mu\nu} .    
\end{equation}
The first and third forms are obtained by solving the second form for $B_{\alpha\beta}$ or $Z_\mu$, respectively. The dual gauge group acts on $B_{\mu\nu}$ as either adding an exact form $\delta B_{\mu\nu}=\partial_{[\mu}C_{\nu]}$ or more generally as adding a closed form $\delta B_{\mu\nu}=\alpha_{\mu\nu}$, $\partial_{[\alpha}\alpha_{\mu\nu]}=0$.

The second loophole is that gauge fixing might discard gauge transformations associated with non-trivial charges. For example, in 3d gravity, Fefferman-Graham gauge in asymptotically anti-de Sitter spacetime discards non-trivial gauge transformations \cite{Grumiller:2017sjh}. In 4d gravity, Bondi gauge and harmonic gauge lead to distinct classes of canonical charges, see e.g. \cite{Compere:2017wrj}. We can resolve these two loopholes by stating that the asymptotic symmetry group of a class of theories is the union of the asymptotic symmetry groups of each formulation of that theory. This enforces that the asymptotic symmetry group is invariant under field redefinitions and gauge choices.

\subsection{Determining the asymptotic symmetry group} \label{subsec:Determing-Asymptotic-Group}

Finding the asymptotic symmetry group of a class of theories is often stated as ``the art of finding consistent boundary conditions''\footnote{This can be attributed at least to Marc Henneaux.}. If one defines too restrictive boundary conditions, some physically important solutions will be discarded. If one defines too large boundary conditions, some important quantities such as the energy will not be defined (but note that many infinities could also be removed using suitable renormalization schemes). In the middle of these uninteresting or unphysical boundaries conditions lies a non-linear zoo of possible interesting boundary conditions. 

In order to give more details, some formalism is required. We will denote the fields of one formulation of the class of theories as $\Phi^i = \{ \phi, A_\mu,g_{\mu\nu},\dots \}$. The Lagrangian density will be denoted as $L[\Phi^i]$. Gauge transformations as $\delta_\lambda \Phi^i = R^i_\lambda [\Phi]$ where the gauge transformation parameters are  $\lambda=\lambda^\alpha(\Phi^i(x^\mu),x^\mu)$. They form an algebra given by $\delta_{\lambda_1}\delta_{\lambda_2}\Phi^i - (1 \leftrightarrow 2)=R^i_{[\lambda_1,\lambda_2]}[\phi]$, $[\lambda_1,\lambda_2]^\alpha= C^\alpha_{\beta\gamma}(\lambda_1^\beta,\lambda_2^\gamma)+\delta_{\lambda_1}\lambda_2^\alpha-\delta_{\lambda_2}\lambda_1^\alpha$. We mainly follow the notation of \cite{Barnich:2001jy,Barnich:2007bf,Compere:2018aar}.

A field symmetry around a given field configuration $\Phi^i$ is a gauge parameter $\lambda^\alpha$ such that its associated gauge transformation vanishes on-shell: $\delta_\lambda \Phi^i=R^i_\lambda[\Phi]=0$. For example, a Killing vector of a spacetime in gravity, or the constant gauge parameter $\lambda=1$ in electromagnetism. Such field symmetries form a group, the exact symmetry group. Exact symmetries only occur for very restricted classes of configurations in interacting theories, e.g. only stationary solutions, or they occur for very restricted classes of theories, e.g. topological theories such as 3d gravity. The group of field symmetries of the maximally symmetric solution is typically used as a benchmark: the asymptotic symmetry group is usually defined such that it contains the exact symmetry group of the maximally symmetric solution \cite{Brown:1986nw} even though they are motivated exceptions \cite{Compere:2013bya}. 

Gauge theories obey the Generalized Noether Theorem for field symmetries \cite{Barnich:1995ap}. Take any physical theory in $n$ spacetime dimensions described by a Lagrangian density which admits field symmetries. It exists a bijection between
\begin{itemize}
    \item the equivalence class of gauge parameters $\lambda(x^\mu)$ that are field symmetries, i.e. such that the variations of all fields $\Phi^i$ vanish on shell $(\delta_\lambda \Phi^i \approx 0)$. Two gauge parameters are equivalent if they are equal on-shell;
    \item The equivalence class of $(n-2)$-forms $\mathbf k$ that are closed on-shell ($d\mathbf k \approx 0$). Two $(n-2)$-forms are equivalent if they differ on-shell by $d\mathbf l$ where $\mathbf l$ is a $(n-3)$-form. 
\end{itemize}
The infinitesimal surface charge $\mathbf k$ can be derived algorithmically from the Lagrangian up to a remaining boundary ambiguity to be fixed by other considerations, see later on. One first defines the presymplectic potential $\Theta$ from the variation of the Lagrangian as \cite{Iyer:1994ys}
\begin{equation}
    \delta L= \frac{\delta L}{\delta \Phi^i}\delta \Phi^i+d\Theta [\delta \Phi^i ; \Phi^i]. 
\end{equation}
All derivatives acting on the fields $\Phi^i$ are integrated by parts in order to define $\Theta$. One then defines the presymplectic structure 
\begin{equation}
    \omega[\delta_1\Phi,\delta_2\Phi ; \Phi]=\delta_1\Theta [\delta_2\Phi ; \Phi]-\delta_2 \Theta[\delta_1\Phi; \Phi]. 
\end{equation}
It is a two-form in field space and $n-1$ form in spacetime. The infinitesimal canonical surface charge is then obtained from the definition 
\begin{equation}
 \omega[\delta_\lambda\Phi,\delta \Phi ; \Phi^i]=dk_\lambda [\delta \Phi ; \Phi] + \text{EOM}
\end{equation}
where the last term on right-hand side is proportional to the equations of motion. An exact field symmetry $\delta_\lambda\Phi^i=0$ leads to a conserved $n-2$ form $dk_\lambda [\delta \Phi ; \Phi]=0$. 

Let us now discuss the remaining boundary ambiguity. One can add a boundary Lagrangian to the action as $S=\int d^{n}x L[\Phi]+\int d^{n-1}x L_B[\Phi ; \Psi]$, which typically depend upon background structures $\Psi$ such as the normal to the boundary. We will denote with a $L$ subscript the bulk Lagrangian contribution and with a $B$ subscript the boundary contribution. This boundary action is associated with the choice of boundary conditions. This boundary action leads to a shift of the presymplectic potential and presymplectic structure by a boundary term as $\Theta = \Theta_L -d \Theta_B$ and $\omega=\omega_L-d\omega_B$ \cite{Papadimitriou:2005ii,Compere:2008us,Compere:2018ylh,Harlow:2019yfa,Freidel:2020xyx}. In turn, this leads to the shift of the infinitesimal surface charge as $k_\lambda=k^L_\lambda[\delta\Phi ; \Phi]-\omega_B[\delta_\lambda \Psi ; \delta \Psi ; \Psi]$. For exact field symmetries we have $\delta_\lambda \Phi^i=\delta_\lambda \Psi^i=0$ and the boundary contribution vanishes. However, for asymptotic symmetries it might contribute, even by an infinite amount in the case where the boundary Lagrangian is chosen to renormalize the infinite variation of the action at the boundary. 

The prime examples of correspondence between exact field symmetries and conserved surface charges are the case of global $U(1)$ symmetry in electromagnetism and Killing vectors in Einstein gravity. The constant $U(1)$ gauge transformation is associated with the electric charge $Q=\int_S \star F[A]$ written in form notation, while the constant dual $U(1)$ gauge transformation is associated with the magnetic monopole charge $P=\int_S F[A]$. In these cases the surface charge one-form $\mathbf k_\lambda$ is an exact variation over the field space, $\mathbf k_\lambda = \delta (\star F \lambda)$ or $\mathbf k_\lambda = \delta (F \lambda)$ where $\lambda=1$. In gravity, the energy $E$ and angular momentum $J$ can be written as 
\begin{equation}
    E,J=\int_S \mathbf K_\xi [g_{\mu\nu}], \qquad \mathbf K_\xi[g_{\mu\nu}]=\int_{\bar g}^{g}\mathbf k_\xi [\delta g_{\mu\nu} ; g_{\mu\nu}]\label{charge}
\end{equation}
where, explicitly, \cite{Iyer:1994ys}
\begin{equation}
   \mathbf K_\xi[g_{\mu\nu}]=\frac{(d^{n-2}x)_{\mu\nu}}{8\pi G} \left[ \sqrt{-g}D^\mu \xi^\nu - \sqrt{-\bar g}\bar D^\mu \xi^\nu + \int_{\bar g}^g \sqrt{-g}\xi^\mu (D^\alpha \delta g^{\nu}_{\alpha}-D^\nu \delta g^\alpha_\alpha) \right]. 
\end{equation}

Let us now discuss asymptotic symmetries at asymptotic or finite boundaries. One can distinguish two categories of boundary conditions: 
\begin{enumerate}
    \item Asymptotically closed boundary conditions are defined as boundary conditions with no energy flux. For example, spatial infinity in asymptotically flat spacetimes by definition cannot be reached by any finite energy \cite{1977asst.conf....1G}. The asymptotic boundary of anti-de Sitter can be reached by finite energy but standard boundary conditions do not allow a flux through the boundary \cite{Ashtekar:1978unified,Brown:1986nw}. Another example is null infinity in three-dimensional Einstein gravity. 
    \item Asymptotically open boundary conditions are defined as boundary conditions with a flux of energy through the boundary (either ingoing or outgoing). The prime example is null infinity in the presence of massless matter or gravity \cite{Bondi:1962px,Sachs:1962aa}. Another example is ``leaky'' anti-de Sitter spacetime which needs to be glued to an exterior geometry \cite{Compere:2020lrt}. 
\end{enumerate}
Let us first discuss the case of closed boundaries. The standard definition of an asymptotic symmetry is as follows. One requires that $\delta \Phi^i=R^i_\lambda[\Phi] \rightarrow 0$ in a suitable way towards the boundary and that the associated surface charge $Q_\lambda$ be finite and conserved. The asymptotic symmetry group is then defined as the quotient \eqref{ASG}. The associated charges then form an algebra \cite{Brown:1986ed,Barnich:2001jy,Barnich:2007bf}
\begin{equation}
    \{ Q_{\lambda_1} , Q_{\lambda_2} \}[\Phi]=Q_{[\lambda_1,\lambda_2]}[\Phi- \bar \Phi]+K_{\lambda_1,\lambda_2}[\bar \Phi]
\end{equation}
under the Peierls bracket \cite{Khavkine:2014kya} where $K_{\lambda_1,\lambda_2}[\bar \Phi]$ is the central extension that only depends upon the reference background $\bar \Phi^i$ with respect to which the surface charges are defined. The requirement of fall-off $\delta \Phi^i=R^i_\lambda[\Phi] \rightarrow 0$ is now often waived in favor of just requiring the charges to be finite and conserved, which is what is ultimately physical. 

In the second case of open boundary conditions, the infinitesimal surface charges $\mathbf k_\lambda[\delta \Phi ; \Phi]$ are not integrable in the sense that $\delta \mathbf k_\lambda[\delta\Phi ; \Phi]\neq 0$. They can be split into an integrable and a non-integrable part as $\delta \mathbf K_\lambda [\Phi]+\Xi_\lambda[\delta \Phi ; \Phi]$ \cite{Barnich:2010eb,Barnich:2011mi} but a prescription is then required to uniquely identify the integrable part with the final surface charge \cite{Wald:1999wa}. This prescription has been related to boundary conditions and associated boundary actions in \cite{Freidel:2020xyx}. The charges $Q_\lambda=\int_S \mathbf K_\lambda$ are required to be finite (allowing a renormalization procedure using counterterms). They also obey an algebra \cite{Barnich:2011mi,Freidel:2021cbc}. We will now turn to the description of explicit cases. 

\section{Celestial asymptotic symmetry groups}

\begin{figure}[htb]
    \centering
    \includegraphics[width = 0.99\linewidth]{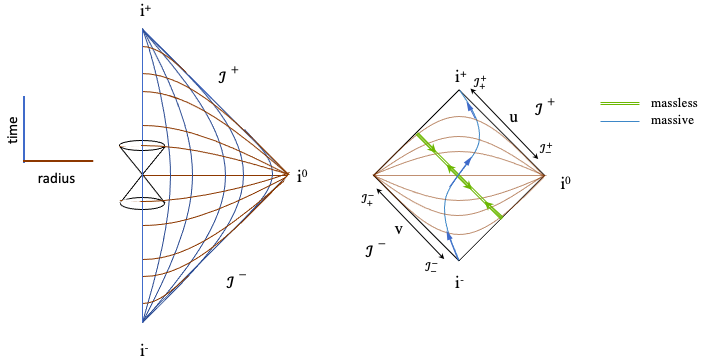}
    \caption{Penrose diagrams of Minkowski spacetime. On the left: Penrose diagram with sample light cone. Vertical curves have constant $t$ while horizontal curves have constant $r$ in standard Minkowski coordinates. The $n-2$ angular dimensions are suppressed. On the right: Massive particles travel from $i^-$ to $i^+$, whereas massless particles travel at $45^o$ incident angle between $\mathcal{I}^-$ and $\mathcal{I}^+$. Advanced ($v = t + r$) and retarded $u = t - r$ coordinates (together with angular coordinates) span the boundaries  $\mathcal{I}^-$ and $\mathcal{I}^+$, respectively.\cite{BSS2021,Pasterski_2019}}
    \label{fig:PM}
\end{figure}

The asymptotic boundary of Minkowski spacetime in $n$ dimensions consists of five parts: future and past null infinities ${\cal I}^\pm$, future and past timelike infinities: Euclidean $AdS_{n-1}$ or ${EAdS}_{n-1}$ (with degenerate $n-2$ sphere at $r=0$ in standard coordinates) usually denoted $i^\pm$, and spacelike infinity: $dS_{n-1}$ or $i^0$ (with smallest sphere at $t=0$ in standard coordinates). Near ${\cal I}^+$ we use retarded coordinates $(u,r,x^A)$  and near ${\cal I}^-$ we use advanced coordinates $(v,r,x^A)$, where $r$ is the radial coordinate, $u = t-r$ is retarded time while $v = t+r$ is advanced time.  For $n=4$, $x^A$ are coordinates on $S^2$, that can be  taken to be the complex stereographic coordinates $(z,\bar{z})$ with $z = e^{i\phi}\tan\frac{\theta}{2}$, $\bar z=z^*$. We define ${\cal I}^\pm$ as the limit as $r \to \infty$ with $(u,x^A)$ or $(v,x^A)$ fixed. In these charts the metric takes the form
\begin{eqnarray}
    ds^2 &=& -du^2-2dudr+r^2\gamma_{AB}dx^Adx^B;\\
    ds^2 &=& -dv^2+2dvdr+r^2\gamma_{AB}dx^Adx^B,
\end{eqnarray}
where $\gamma_{AB}$ is the $S^{n-2}$ round metric. Following \cite{Strominger:2013jfa}, we also define the boundaries of ${\cal I}^+$ to be the spheres ${\cal I}^+_\pm$ defined as the limit $u\to \pm \infty$ taken after $r\rightarrow \infty$ and similarly we define the boundaries of ${\cal I}^-$ to be ${\cal I}^-_\pm$ defined as the limit $v\to \pm \infty$  taken after $r\rightarrow \infty$, see Figure \ref{fig:PM}.

The retarded and advanced coordinates are not appropriate to describe $i^0$ and $i^\pm$. For these boundaries we employ a hyperbolic slicing of Minkowski spacetime \cite{Ashtekar:1978unified,Campiglia:2015yka}. Near $i^0$ we introduce coordinates $(\tau,\rho,x^A)$ where $t =\rho\sinh\tau$ and $r = R+\rho\cosh\tau$, for large $R$, with $i^0$ being defined as the $\rho \to \infty$ limit. Near $i^\pm$ we introduce coordinates $(\hat{\tau},\hat{\rho},x^A)$ where $t = \pm T+\hat{\tau}\cosh\hat{\rho}$ and $r = \hat{\tau}\sinh\hat{\rho}$, for large $T$, and then $i^\pm$ are defined as the surfaces $\hat{\tau} \to \pm \infty$. In these two charts the metric reads, respectively, as
\begin{eqnarray}
    ds^2 &=& d\rho^2 +\rho^2\left(-d\tau^2+\cosh^2\tau\,  \gamma_{AB}dx^Adx^B\right)+O(\rho),\\
    ds^2 &=& -d\hat{\tau}^2 +\hat{\tau}^2\left(d\hat{\rho}^2+\sinh^2\hat{\rho}\, \gamma_{AB}dx^Adx^B\right)+O(\hat \tau).
\end{eqnarray}

Let us end with a general remark on the space of fields in Minkowski spacetime. In Minkowski spacetime in spacetime dimension $d \geq 4$, the infrared structure of any propagating field at each asymptotic boundary can be decomposed in terms of multipole moments which appear at higher and higher subleading orders in the expansion from each boundary. In that sense, there is an infinite amount of holographic fields dual to a propagating field in Minkowski spacetime. This is very different than in anti-de Sitter spacetime where only two holographic fields (the ``source field'' and its ``vacuum expectation value'') appear in the radial expansion from the boundary \cite{Witten:1998qj}.

\subsection{$4d$ QED - $U(1)$ Kac-Moody}

We now consider QED by first looking only at the gauge potential $A_\mu$. We work in the gauge fixing approach and gauge fix to radial gauge and boundary temporal gauge:
\begin{eqnarray}
    {A}_r = 0, \qquad  {A}_u|_{\cal I^+} = 0.
\end{eqnarray}
A gauge transformation has the form $\delta_\lambda {A}_\mu = e^2\partial_\mu\lambda$ and it preserves radial gauge when $\lambda = \lambda(x^A)$. In solution space, at ${\cal I}^+_\pm$, the field configuration is pure gauge $A_A^\pm(x^B) = e^2\partial_A\phi^+_\pm(x^B)$ and the gauge invariant difference $\phi^+_+(x^A)-\phi^+_-(x^A)$ characterizes the electric memory effect \cite{Bieri:2013em,Pasterski:2017asem}. The only field left undetermined from $F_{\mu\nu}$ is $\phi^+_-(x^A)$ and it transforms inhomogeneously under non-trivial gauge transformations, $\delta_\lambda \phi^+_-(x^A) = \lambda(x^A)$, which allows its identification as the Goldstone boson of the spontaneously broken large gauge symmetry \cite{He:2014new}. There is a dual analogue to this construction. 

After using the definition of the canonical charge and the field equations, the Noether charge $Q_\lambda^+$ associated to the gauge transformation with parameter $\lambda$ is
\begin{eqnarray}\label{eq:large-gauge-charge-massless-qed}
    Q_\lambda^+ &=& \dfrac{1}{e^2}\int_{\cal I^+_-}\sqrt{\gamma}d^2z \lambda F_{ru}\\ &=&\underbrace{\int_{S^2}\sqrt{\gamma}d^2z \lambda D^CD_C(\phi^+_+-\phi^+_-)}_{\text{Soft/Memory Part}}+\underbrace{\int_{\cal I^+}du\sqrt{\gamma}d^2z\lambda j_u}_{\text{Hard Part}},
\end{eqnarray}
where $j_u$ is a possible massless ${\rm U}(1)$ current sourced by charged matter. Since this contribution admits a finite charge it is called the ``hard'' part. The first term has not monopole charge because when $\lambda=1$ the integrand is a total derivative and the integral vanishes from Stokes' theorem. Yet, there is a non-trivial contribution for arbitrary $\lambda$ which is sourced by $\phi^+_+-\phi^+_-$, also sourcing the electric memory effect.  For this reason it is called the soft or memory part. The associated asymptotic symmetry algebra is abelian $\{ Q_\lambda^+,Q_{\lambda'}^+ \}=0$. 

We can do the same analysis on ${\cal I}^-$. It relates to the one we have outlined above at ${\cal I}^+$ after imposing junction conditions at spatial infinity. One requires that $\phi^+_-(x^A)$ be antipodally matched to the corresponding $\phi^-_+(x^A)$ and that the asymptotic symmetries preserve this matching. The antipodal matching is necessary for the scattering problem to be well-defined and Lorentz invariant \cite{He:2014new} and it has been verified in the Li\'enard-Wiechert solution \cite{Strominger:2017zoo}. It can also be understood directly from a Hamiltonian analysis from twisted parity conditions at the constant $t$ surface at $r\rightarrow \infty$ (also known as spatial infinity $i^0$) \cite{Henneaux:2018asymptotic} or from a Lagrangian analysis from the behavior of waves on the boundary $dS_3$ \cite{Prabhu:2018conservation} .

We illustrate the inclusion of massive fields with a massive charged scalar $\varphi$. The field equations are $\nabla^\nu F_{\mu\nu}= J_\mu$ and $(-D_\mu D^\mu +m^2)\varphi=0$ where $J_\mu=i e \varphi (D_\mu \varphi)^* + c.c.$ and $D_\mu\varphi=\partial_\mu \varphi -i e A_\mu \varphi$. In this case the relevant Cauchy surface is $\Sigma_+={\cal I}^+\cup\,  i^+$, with $i^+$ represented as Euclidean $AdS_3$ via the hyperbolic slicing, and with $\partial i^+$ matched with ${\cal I}^+_+$ locally at each angle $x^A$ \cite{Campiglia:2015yka}. In order to have consistency between ${\cal I}^+$ and $i^+$ radial gauge is not convenient and we instead gauge fix to Lorenz gauge $\nabla^\mu A_\mu=0$. Residual gauge transformations $\delta_{\tilde{\lambda}} A_\mu = e^2\partial_\mu\tilde{\lambda}$ and $\delta_{\tilde{\lambda}}\varphi = i\tilde{\lambda}\varphi$ are now constrained by $\Box\tilde{\lambda}=0$.

At ${\cal I}^+$ the gauge parameter $\tilde{\lambda}$ still asymptotes to a function $\lambda(x^A)$ of the angles while at $i^+$ it asymptotes to a function $\lambda_{\cal H}(\rho,x^A)$ on ${EAdS}_3$ constrained, by the residual gauge condition $\Box \tilde{\lambda} =0$ and by compatibility with the value of $\tilde{\lambda}$ at ${\cal I}^+$, to be a solution of the problem 
\begin{eqnarray}
    \Delta\lambda_{\cal H} = 0,\quad \lim_{\rho\to \infty}\lambda_{\cal H}(\rho,x^A) = \lambda(x^A),
\end{eqnarray}
where $\Delta$ is the Laplacian on ${EAdS}_3$ and its coordinates are defined with $\rho\to \infty$ reaching $\partial i^+$ \cite{Campiglia:2015yka}. This problem is solved by a scalar bulk-to-boundary propagator applied to $\lambda(x^A)$ in a manner analogue to the $AdS_3$ propagator  \cite{Witten:1998qj,Freedman:1998tz}.

The Noether charge that follows from the covariant phase space procedure still has the same expression given in Eq. (\ref{eq:large-gauge-charge-massless-qed}) as a surface integral over $\partial \Sigma_+ = {\cal I}^+_-$. However, when rewritten as a codimension 1 integral over ${\Sigma_+}$ we have a new contribution from the massive field at $i^+$: \cite{Campiglia:2015yka,Kapec:2015ena}
\begin{eqnarray}
    Q_\lambda^+ &=&\underbrace{\int_{S^2}\sqrt{\gamma}d^2z \lambda D^CD_C(\phi_+-\phi_-)}_{\text{Soft/Memory Part}}+\underbrace{\int_{\cal I^+}du\sqrt{\gamma}d^2z\lambda j_u}_{\text{Hard Massless Part}}+\underbrace{\int_{ i^+}d\rho d\Omega\sqrt{q}\lambda_{\cal H} j_\tau}_{\text{Hard Massive Part}},
\end{eqnarray}
where ${\lambda}_{\cal H}$ is implicitly written in terms of $\lambda(x^A)$ through the bulk-to-boundary propagator \cite{Campiglia:2015yka, Strominger:2017zoo}.

We close this discussion with a comment on an alternative set of asymptotic symmetries in QED known as \textit{multipole symmetries}. In this case the gauge parameter is of the form $\lambda_{\ell m}(r,x^A) = r^\ell Y_{\ell m}(x^A)$ and the surface charges $Q_{\ell,m}^+$ for an electrostatic field are the electric multipole moments  \cite{Seraj:2017multipole}. The asymptotic symmetry algebra is again a $U(1)$ Kac-Moody algebra. 

\subsection{$3d$ flat gravity - BMS$_3$}

Three-dimensional asymptotically flat gravity is trivial in the sense that there is no propagating degree of freedom and no black hole, according to Ida's theorem \cite{Ida:2000jh}. Yet, it is non-trivial in the sense of boundary dynamics or asymptotic symmetries: one can still define boundary conditions such that the asymptotic symmetry group is non-trivial. In addition, while there is no Newtonian attractive potential in $3d$ gravity, there are still spinning massive particles that are defined as conical defects, with a time twist in the presence of angular momentum. 

Let us review this asymptotic structure in the gauge fixing approach and in the Lagrangian formalism for the metric field $g_{\mu\nu}$. The action is the Einstein-Hilbert action. We will first define the space of solutions at null infinity in Newmann-Unti gauge or, equivalently, as we will show, in Bondi gauge, and derive the asymptotic symmetry charge algebra for given boundary conditions. We will then investigate spatial infinity and derive boundary conditions that are consistent with the antipodal identification of the symmetry groups between $\mathcal I^+$ and $\mathcal I^-$.

We denote the spacetime coordinates as $(u,r,\phi)$. In the following, $\mu, \nu$ will stand for spacetime indices while $A, B$ will stand for sphere indices. We impose the so-called  Newman-Unti gauge condition
\begin{align}
g_{rr} = g_{r\phi} = 0,\qquad g_{ru} = -1. \label{NU3d}
\end{align}
As we will see in a moment, the imposition of boundary conditions will result in $g_{\phi\phi}=r^2$ and the radial coordinate will also equal the luminosity distance in the sense of \cite{Sachs:1962wk} and the gauge fixing will be equivalent to Bondi gauge. 

Consistently with the inclusion of Minkowski spacetime and spinning conical defects, we impose the following boundary conditions \cite{Barnich:2006av}
\begin{equation}
g_{uu} = \cO(r^0), \qquad g_{uA} = \cO(r^0), \qquad g_{\phi \phi} = O(r). \label{bc3d}
\end{equation}
We will need to check whether the conserved quantities will be finite given these boundary conditions.
  Given this gauge choice and boundary conditions, the most general form of a metric solving Einstein's equation can be written as follows \cite{Barnich:2010eb,Barnich:2012aw}
\begin{equation}
ds^2 = \Theta(\phi) du^2 - 2 du dr + 2 \left[ \Xi(\phi) + \frac{u}{2} \p_{\phi} \Theta(\phi) \right] du d\phi + r^2 d\phi^2.\label{3dmetric}
\end{equation}
The phase space of the theory is then the set of all such metrics parameterized by the 2 functions $\Theta$ and $\Xi$ on the circle. The residual symmetries preserving the gauge choice and boundary conditions can be deduced from the condition
\begin{equation}
\cL_{\xi} g_{\mu \nu} (\Theta,\Xi) = g_{\mu \nu}(\Theta + \d_{\xi}\Theta, \Xi + \d_{\xi} \Xi) - g_{\mu \nu} (\Theta,\Xi), \label{lie3}
\end{equation}
which is imposed at linear order in the vector $\xi^\mu$. It implies in particular $\mathcal L_\xi g_{r\mu}=0$. The general solution to Eq. \eqref{lie3} is
then 
\begin{equation}\label{allx}
\xi^{\mu}\p_{\mu} : \begin{cases}
\xi^{u} = T(\phi) + u \p_{\phi} R(\phi); \\
\xi^r = - r \p_{\phi} R(\phi) + \p_{\phi}^2 T + u \p_{\phi}^3 R(\phi)-\frac{1}{r}(\partial_\phi T + u \partial_\phi^2 R)(\Xi+\frac{u}{2}\partial_\phi \Theta) ; \\
\xi^{\phi} = R(\phi) - \frac{1}{r} \p_{\phi} T - \frac{u}{r} \p_{\phi}^2 R(\phi).
\end{cases}
\end{equation}
Here, the supertranslations are generated by $T$ and superrotations by $R$. If we define the modes by,
\begin{align}
P_m = \xi (T = e^{im\phi}, R=0) ,\qquad 
J_m = \xi (T=0, R = e^{im\phi}),
\end{align}
it is easy to see that we get the following asymptotic algebra under the standard Lie bracket \cite{Ashtekar:1996cd}
\begin{align}
i \left[ P_m, P_n \right] &= 0, \\ i \left[ J_m, J_n \right] &= (m-n) J_{m+n}, \\ i \left[ J_m, P_n \right] &= (m-n) P_{m+n}.
\end{align}
In fact, the algebra can even be promoted to be exact at any $r$ given the exact form \eqref{allx} but the Lie bracket $[\xi,\eta]$ needs to be enhanced to the adjusted Lie bracket $[\xi,\eta]_*=[\xi,\eta]-\delta_\xi \eta+\delta_\eta \xi$ which takes into account the field dependence (here in $\Xi(\phi)$, $\Theta(\phi)$) of the vector fields \cite{Barnich:2010eb}. 

The action of the vector fields can be understood as transformations directly on the solution space using \eqref{lie3} from which we get,
\begin{align} 
\d_{T,R} \Theta &= R \p_{\phi} \Theta + 2 \Theta \p_{\phi}R - 2 \p_{\phi}^3 R, \label{theta_trans}\\
\d_{T,R} \Xi &= R \p_{\phi} \Xi + 2 \Xi \p_{\phi} R + \frac{1}{2} T \p_{\phi} \Theta + \Theta \p_{\phi} T  - \p_{\phi}^3 T. \label{Xi_trans}
\end{align}
By virtue of \eqref{theta_trans}, $\Theta$ belongs to the coadjoint representation of Diff$(S^1)$.  The above equations simplify when written in terms of simpler fields. With hindsight \cite{Barnich:2014kra}, we define the superrotation field $\Psi (\phi)$, which is invariant under supertranslations and transforms under superrotations as,
\begin{equation}\label{dPsi}
\delta_{T,R} \Psi = R \partial_{\phi} \Psi + \partial_{\phi} R. 
\end{equation}
This implies that $\Theta = \left( \partial_{\phi} \Psi \right)^2 - 2 \partial^2_{\phi} \Psi + 8GM e^{2 \Psi}$, which gives it the form of a Liouville stress tensor. The parameter $M$ is exactly the charge conjugate to $P_0 = \partial_t$. For $\Psi =0$, we have $\Theta_0 = 8 G M$ and $M$, which corresponds to the total conical defect, cannot be generated by a diffeomorphism. Similarly, for the other transformation, we can introduce a supertranslation field $C$, from which we define $\Xi$ as
\begin{equation}
\Xi = \Theta \partial_{\phi} C - \partial^3_{\phi} C + 4 G J e^{2 \Psi} + \frac{1}{2} \partial_{\phi} \Theta C.
\end{equation}
This along with the previous equation fixes the transformation of the $C$ field as,
\begin{equation}\label{dC}
\delta_{T,R} C = T + R\partial_{\phi}C - C \partial_{\phi} R.
\end{equation}
Like $M$, the zero mode $\Xi_0 = 4G J$ is recognized after computing the charges as determined by the angular momentum $J$ conjugate to $- \partial_{\phi}$. As a summary, the field space is now parameterized by the supertranslation field $C(\phi)$, the superrotation field $\Psi(\phi)$, and the zero modes $J$ and $M$. 

Given the symmetry algebra, we can find the corresponding generalized Noether charges and compute the corresponding algebra. The expression of the charge \eqref{charge} with convention $\epsilon_{ur\phi}=1$ gives 
\begin{align}
\mathcal{P}_n &= \frac{1}{16 \pi G} \int_0^{2 \pi} d \phi \left( \Theta(\phi) + 1 \right) e^{i n \phi}, \\
\mathcal{J}_n &=  \frac{1}{8 \pi G} \int_0^{2 \pi} d \phi \, \Xi(\phi) \, e^{i n \phi}.
\end{align}
Using the bracket algebra defined as $\left[ Q_{\xi}, Q_{\xi'} \right] = \delta_{\xi'} Q_{\xi}$ and the transformation laws \eqref{dPsi}-\eqref{dC}, we get the following charge algebra \cite{Barnich:2006av}
\begin{align}
i \left[ \mathcal{P}_m, \mathcal{P}_n \right] &= 0, \nonumber\\ 
i \left[ \mathcal{J}_m, \mathcal{J}_n \right] &= (m-n) \mathcal{J}_{m+n}, \label{BMS3}\\
i \left[ \mathcal{J}_m, \mathcal{P}_n \right] &= (m-n) \mathcal{P}_{m+n} + \frac{1}{4 G} m (m^2 -1) \delta_{m+n,0},\nonumber
\end{align}
where we now have a central extension.  This is due to the third derivative terms in the transformation of $\Theta$ and $\Xi$ in \eqref{theta_trans}, \eqref{Xi_trans}. In 3 dimensions, the angular momentum is dimensionless while the momentum has dimension inverse length. The central charge has therefore dimension inverse length. Only the generators corresponding to the exact Killing generators of the Poincar\'e algebra $\mathfrak{iso}(2,1)$ generated by $\mathcal{P}_0,\mathcal{P}_{-1}, \mathcal{P}_1$, $\mathcal{J}_{-1}, \mathcal{J}_0, \mathcal{J}_1$ do not have a central extension. In the quantum theory, $i$ times the Peierls bracket $i[\_,\_]$ is the quantum commutator. The above algebra can also be obtained from a In\"on\"u-Wigner contraction of the Virasoro $\times$ Virasoro algebra \cite{Brown:1986nw} that arises as the symmetry group for asymptotically anti-de Sitter spacetimes \cite{Barnich:2006av} .

To complete the story, we investigate the asymptotic region of spatial infinity. While there is considerable physical justification for the antipodal map that relates the BMS group BMS$^+$ at $\cI^+$ to the BMS group BMS$^-$ at $\cI^-$ \cite{Strominger:2013jfa}, it is interesting to see whether one can derive this map from a fundamental perspective. Spatial infinity is the place to look at  since it is bounded by $\cI^+_-$ the past boundary of $\cI^+$ and $\cI^-_+$ the future boundary of $\cI^-$. Hyperbolic gauge is reached using the gauge fixing conditions 
\begin{equation}
g_{\rho \rho} = 1, \quad g_{\rho \tau} = g_{\rho \phi} = 0.
\end{equation}
The boundary conditions specifying the set of asymptotically flat spacetimes at spatial infinity are
\begin{equation}
g_{\tau \tau} = O(\rho^2), \quad g_{\tau \phi} = O (\rho^2), \quad g_{\phi \phi} = O(\rho^2).
\end{equation}
In particular, the Minkowski metric is $ds^2 = d\rho^2 + \rho^2 \left( - d\tau^2 + \cosh^2 \tau \, d \phi^2 \right)$, with the usual Minkowski time and radius is given by $t = \rho \sinh \tau, \, r = \rho \cosh \tau$. The general exact solution in this gauge can be written down as
\begin{equation}
ds^2 = d \rho^2 + \left( \rho^2 h^{(0)}_{ab} + \rho h^{(1)}_{ab} + h^{(2)}_{ab} \right) dx^a dx^b,
\end{equation}
where the entire metric can be reconstructed from two holographic ingredients (this is specific to 3d gravity!): the boundary metric $h^{(0)}_{ab}$ and the boundary stress-tensor $T^{ab}$ ($a,b$ running over boundary coordinates) as
\begin{align}
h^{(1)}_{ab} &= T_{ab} - h^{(0)}_{ab} h^{(0)}_{cd} T^{cd} , \qquad  h^{(2)}_{ab} = \frac{1}{4} h^{(1)}_{ac} h_{(0)}^{cd} h^{(1)}_{db}.
\end{align}
Einstein’s equations imply that the boundary metric is locally $dS_2$ with boundary Ricci scalar $R_{(0)}=2$, and that the stress-tensor is conserved $\mathcal{D}_aT^{ab} = 0$. Here $\mathcal{D}_a$ is the boundary covariant derivative with respect to $h^{(0)}_{ab}$ and indices are raised with the inverse
boundary metric $h^{ab}_{(0)}$. The trace of the stress-tensor is not fixed, contrary to the analogous Fefferman-Graham expansion in $AdS_3$ \cite{deHaro:2000vlm}. 

To have a well-defined variational principle (and obtain integrable charges) however, one needs to impose additional boundary conditions which are chosen as  \cite{Compere:2017knf} 
\begin{equation}
h^{(0)}_{++} = h^{(0)}_{--} = 0, \qquad T_{+-} = 0.
\end{equation}
where we defined the boundary lightcone coordinates $x^{\pm} = \tau \pm \phi$. The most general boundary metric is then given by,
\begin{equation}
ds^2_{(0)} = - \frac{2 \Delta^2 \partial_+ X^+ \partial_- X^-}{\cos \left(\Delta (X^+ + X^-)\right) + 1} dx^+ dx^-,
\end{equation}
where $X^+ = X^+(x^+)$ and $X^- = X^-(x^-)$ are chiral functions and $\Delta$ is the conical defect in the boundary de Sitter space. We also have that $0 < \Delta \leq 1$ where $\Delta=1$ is equivalent to the absence of defect. After computing the change of coordinates to Bondi gauge, one observes that preserving asymptotic flatness at the future and past null infinities in the sense of \eqref{bc3d} requires the following identification of the fields
\begin{equation}
X^- (x) = \frac{\pi}{\Delta} - X^+ \left( \frac{\pi}{\Delta} - x \right) + \frac{2 \pi}{\Delta}k, \qquad \forall x,
\end{equation}
where $k \in \mathbb{Z}$ labels disjoint BMS orbits.  The metric is then given by 
\begin{equation}
ds^2=d\rho^2+2h_{+-}^{(0)}\left(\rho dx^+ + \frac{\Xi_-(x^-)}{2h_{+-}^{(0)}}dx^- \right)   \left(\rho dx^- + \frac{\Xi_+(x^+)}{2h_{+-}^{(0)}}dx^+ \right)   . 
\end{equation}
At this point the phase space is characterised by 3 independent null boundary fields $X^+(x^+)$, $\Xi_+(x^+)$, $\Xi_-(x_-)$. However only a particular combination of $\Xi_\pm$ appears in the charges which implies that $\Xi_-$ can be considered pure gauge. The fields at spatial infinity are finally related to the fields on null infinity in Bondi gauge in the following way,
\begin{align}
\Xi (\phi) &= \frac{1}{2} \left[ \Xi_+ \left(\frac{\pi}{2 \Delta} + \phi \right) -  \Xi_- \left(\frac{\pi}{2 \Delta} - \phi \right) \right], \\ 
\Theta(\phi) &= \left( \partial_{\phi} \Psi (\phi) \right)^2 - 2 \partial^2_{\phi} \Psi (\phi) - \Delta^2 e^{2 \Psi (\phi)},
\end{align}
where $e^{\Psi (\phi)} \equiv \partial_{\phi} X^+ \left( \frac{\pi}{2 \Delta} + \phi \right)$. The BMS$_3$ algebra is therefore realised at spatial infinity. The antipodal map between $\mathcal I^+_-$ and $\mathcal I^-_+$ in the absence of defect, $\Delta=1$, follows from the fact that null fields such as $X^+(x^+)$, $\Xi_\pm(x^\pm)$ propagate on $dS_2$ from $\phi$ to $\phi+\pi$ from the past to future of $dS_2$ \cite{Compere:2017knf}.

\subsection{$4d$ flat gravity - BMS$_4$ and more}
\label{BMS4}

The motivation into to study asymptotically flat spacetimes in $3+1$ dimensions is obvious. It describes the physics of localized objects and events in gravity below cosmological scales. Many approaches have been developed over the years, especially since the sixties, to study such classes of spacetimes and many recent developments took place following in particular the seminal works of Barnich-Troessaert \cite{Barnich:2009se} and Strominger \cite{Strominger:2013jfa}. 

Contrary to 3 dimensions, asymptotically flat spacetimes in 4 dimensions admit 2 polarisation modes of gravitational waves, attractive gravitational potentials and black holes. Such classes of spacetimes cannot be written in an exact form and expansions close to boundaries are required. We will follow here the gauge fixing approach within the Lagrangian formalism for the metric field. 

Let us first consider future null infinity. We introduce retarded spherical coordinates $(u,r,\theta,\phi)$ and impose the following boundary conditions,
\begin{align}
&g_{uu} = \mathcal{O}(r^0), \qquad g_{AB} = \mathcal{O}(r^2), \qquad
\sqrt{\det(g_{AB})} = r^2 \sqrt{\det(\bar{q}_{AB})} + \mathcal{O}(r^1), \label{eq:bc_bms4}
\end{align}
where $\bar{q}_{AB}$ is the unit sphere metric. The last condition on the determinant can be further relaxed, which leads to additional Weyl asymptotic symmetries \cite{Freidel:2021fxf}, but we shall not consider such an extension here. Consistently with Einstein's equations, a diffeomorphism exists \cite{Compere:2018ylh} such that $g_{AB}=r^2 \bar q_{AB}+O(r)$ which is the standard asymptotically flat coordinate system. However, it is instructive to consider the more general set of coordinates \eqref{eq:bc_bms4} precisely because the diffeomorphism reducing the metric to standard form is non-trivial in the sense that it is associated with non-vanishing charges and flux-balance laws. 

Assuming these boundary conditions and \emph{past stationarity}, i.e. the absence of any gravitational radiation past a fixed retarded time $u=-T$, future null infinity is asymptotically simple in the sense of Penrose \cite{Penrose:1962ij}, and can be charted by a large class of coordinate systems that admit an expansion in terms of inverse radial powers \cite{Blanchet:1986dk}. We will work with the following gauge choice,
\begin{equation}
g_{rr} = 0 = g_{rA},\qquad  \quad g_{ru} = -1,
\end{equation}
which is called Newman-Unti gauge. This is equivalent \cite{Barnich:2011ty} to another gauge choice that is commonly used, called Bondi gauge where the last condition $g_{ru}=-1$ is replaced by,
\begin{equation}
\partial_r \left( \frac{\text{det}(g_{ab})}{r^4} \right) = 0.
\end{equation}
The asymptotic solution to Einstein's equation with these boundary conditions and this gauge choice is then given by \cite{Bondi:1962px,Sachs:1962wk,Barnich:2010eb} 
\begin{align}
ds^2 &= - \frac{\mathring{R}}{2} du^2 - 2 du dr + r^2 q_{AB}dx^A dx^B \\
& \quad + \frac{2m}{r} du^2 + r C_{AB} dx^A dx^B + \dots \\
&\quad + \frac{1}{r} \frac{4}{3} N_A du dx^A + \dots
\end{align}
where $\mathring{R}$ is the $2$d curvature associated to the metric on the $2$-sphere $q_{AB}$, $m$ is called the Bondi mass aspect, and the $N_A$ is called the angular momentum aspect. The $C_{AB}$ is a symmetric traceless tensor called the \emph{shear},  which contains the 2 polarization modes of the gravitational waves at leading order in the asymptotically flat region at null infinity. It is not fixed by Einstein's equations in the expansion around null infinity and is rather fixed by the source producing the radiation \cite{Thorne:1980ru,Blanchet:1985sp,Flanagan:2015pxa,Blanchet:2020ngx}. The Bondi news is given by $N_{AB} = \frac{d}{du} C_{AB} = \dot{C}_{AB}$. Any metric of the sphere $q_{AB}$ can be written as a Weyl transformation combined with a 2-diffeomorphism $x^a \mapsto G^a(x^b)$ applied on the complex plane metric $\gamma_{ab}$ as $q_{AB}=e^{-\Phi}\partial_A G^a \partial_B G^b \gamma_{ab}$.  Imposing the last boundary condition \eqref{eq:bc_bms4} fixes $\Phi$ in terms of $\text{det}(\bar q_{AB})$ and $G^a$. We call $\Phi$ the superboost field. The 2d Ricci scalar of $q_{AB}$ is then given by $\mathring R=D^A D_A \Phi$ where $D_A$ is the covariant derivative compatible with the metric $q_{AB}$. For a 2d metric, the analogue of the Weyl tensor is the trace-free part of the Liouville stress-tensor built from $\Phi$ that we name $N_{AB}^{\text{vac}}=[\frac{1}{2}D_A \Phi D_B\Phi -D_A D_B \Phi]^{TF}$ and which corresponds up to a trace to the Geroch tensor \cite{1977asst.conf....1G}. In standard coordinates where $q_{AB}=\bar q_{AB}$, this tensor vanishes: $N_{AB}^{\text{vac}}=0$.

The residual gauge transformations that obey $\mathcal{L}_{\xi} g_{r\mu} = 0$ and the boundary conditions \eqref{eq:bc_bms4} are parameterized by 3 functions on the celestial sphere,
\begin{equation}
T(\theta,\phi), \quad R^A(\theta,\phi), \qquad A = \theta, \phi.
\end{equation}
Here $T$ corresponds to \emph{supertranslations} and $R^A$ is associated to \emph{super-Lorentz}\footnote{In most of the literature these are called \emph{superrotations}. We find however convenient to denote the divergence-free ($D_A R^A=0$) generators as superrotations and the curl-free generators ($\epsilon^{AB}\partial_A R_B=0$) as superboosts which generalize the rotations and boosts, respectively.} transformations. Since the charges associated with these residual transformations are finite (after necessarily non-covariant \cite{Flanagan:2019vbl} renormalization \cite{Compere:2018ylh} in the case of super-Lorentz charges), all residual transformations are asymptotic symmetries \cite{Barnich:2009se,Barnich:2010eb,Barnich:2011mi,Campiglia:2015yka}. In the following we will consider the classical asymptotic symmetry group where all functions are regular over the sphere such that the associated charges are finite. Punctures are necessary in the quantized theory and poles in stereographic coordinates $z=\cot\frac{\theta}{2}e^{i \phi}$ and $\bar z=\cot\frac{\theta}{2}e^{-i \phi}$ are then allowed\footnote{The restriction to holomorphic and anti-holomorphic functions for the super-Lorentz transformations is often considered, as in the original work \cite{Barnich:2009se} or in \cite{Kapec:2014opa}. Non-holomorphic functions are required for the correspondence between subleading soft theorems and super-Lorentz transformations \cite{Campiglia:2014yka}. Relationships between $\text{Diff}(S^2)$ and the Virasoro$\times$Virasoro algebra have been established using so-called shadow transforms \cite{Pasterski:2021fjn}. The higher dimensional generalization to $\text{Diff}(S^{n-2})$ is natural while there is no known generalization of the group of holomorphic and anti-holomorphic super-Lorentz transformations.}.

The supertranslation vector fields are given by,
\begin{equation}\label{xiT}
\xi(T)\equiv T(\theta,\phi) \partial_u + \frac{1}{2} D^A D_A T \partial_r - \frac{1}{r} \partial^A T \partial_A + \dots ,
\end{equation}
where dots indicate subleading terms in $r$ that enforce the gauge fixing conditions. These vector fields form an abelian ideal of the full BMS\textsubscript{4} algebra to be detailed below. Their associated Noether charge is the Bondi mass aspect with a correction, $\overline{M}(u,x^A)=m(u,x^A)+\frac{1}{8}C_{AB}(u,x^A) N^{AB}_{\text{vac}}(x^A)$ \cite{Bondi:1962px,Sachs:1962aa}
\begin{equation}
Q_T(u) = \int_{S^2} d^2 \Omega \, T (x^A) \overline{M} (u,x^A),
\end{equation}
where the measure over the sphere is $d^2\Omega=\sqrt{\text{det}(\bar q_{AB})}d^2x$. The four lowest harmonics of $T$ correspond to the four translations, which are associated with the four lowest harmonics of $m$: the $l=0$ harmonic corresponds to a time translation, while the $l=1$ modes correspond to spatial translations. For Minkowski, the constraint equations on $m$ and $N_A$ obtained from Einstein's equations fix $\overline{M}=0$. 

The (super-)Lorentz vector fields are given by,
\begin{equation}    \label{xiR}
\begin{split}
\xi(R)\equiv \left( R^A - \frac{u}{2r} D^A D_B R^B + \mathcal{O} \left(\frac{1}{r^2}\right)\right) \partial_A + \frac{1}{2} u D_A R^A \partial_u &+ \left(-\frac{1}{2}(r+u)D_A R^A + \mathcal{O}\left(\frac{1}{r}\right)\right) \partial_r ,
\end{split}
\end{equation}
where $R^A$ in general can be decomposed as $R^A = \epsilon^{AB} \partial_B \phi + \partial^A \psi$, with the first term corresponding to (super)rotations and the second term corresponding to (super)boosts. After renormalization and a choice of prescription, their associated Noether charge is the Bondi angular momentum aspect $N_A (u,x^A)$ plus a given correction:
\begin{equation}
Q_R(u) = \int_{S^2} d^2 \Omega R^A(x^A) ~ \overline{N}_A (u,x^A),
\end{equation}
where $\overline{N}_A = N_A+ \dots$. There are several conventions for defining $N_A$ from the metric, and several nonequivalent prescriptions for defining the correction to $N_A$ in the literature at fixed finite $u$, see \cite{Barnich:2011mi,Flanagan:2015pxa,Hawking:2016sgy,Compere:2019gft,Campiglia:2020qvc,Compere:2020lrt,Freidel:2021fxf}. The 6 lowest harmonics give the usual Lorentz charges - the angular momenta and center-of-mass.

The (generalized) BMS algebra of the vector fields \eqref{xiT}-\eqref{xiR} reads as 
\begin{equation}    \label{alg}
\begin{split}
&[\xi(T_1),\xi(T_2)]_*=0, \\
&[\xi(R_1),\xi(T_2)]_*=\xi(\hat T),\qquad \hat T=R_1^A \partial_A T_2-\frac{1}{2}D_A R^A_1 T_2 , \\
&[\xi(R_1),\xi(R_2)]_* = \xi(\hat R),\qquad \hat R^A=R_1^B \partial_B R^A_2 - R_2^B \partial_B R_1^A, 
\end{split}
\end{equation}
where $[\xi,\eta]_*=[\xi,\eta]-\delta_\xi \eta+\delta_\eta \xi$ is the adjusted Lie bracket \cite{Barnich:2010eb}. This BMS algebra is represented at the level of the charges $Q_\xi$ but with a non-standard Peierls bracket due to the non-conservation of these charges along $u$ \cite{Barnich:2011mi,Distler:2018rwu,Compere:2020lrt,Freidel:2021cbc}. In particular, the bracket $\{ Q_{T=\partial_u}, Q_\xi\}$ is equivalent to the flux-balance laws that dictate the $u$ evolution of the Bondi mass and angular momentum aspects \cite{Barnich:2011ty}, which follow from Einstein's equations in Bondi or Newmann-Unti gauge.

What we have seen so far is the symmetry at $\mathcal{I}^+$ and it is clear that there is a similar story at $\mathcal{I}^-$ when analysed in advanced coordinates $(v,r,\theta,\phi)$. In order to define scattering from $\mathcal I^-$ to $\mathcal I^+$, it is necessary to define \emph{junction} conditions between the fields at $\mathcal I^-$ and $\mathcal I^+$ at spatial infinity. Strominger proposed to relate antipodally (on the 2-sphere) the charges at $\mathcal{I}^{+}_{-}$ and the charges at $\mathcal{I}^{-}_{+}$ in accordance with CPT and Lorentz invariance \cite{Strominger:2013jfa}. For the charges with lowest harmonics this can also be justified from consistency with the boosted Kerr black hole\footnote{[Strominger, private communication]}. For the restricted BMS group without super-Lorentz transformations, boundary conditions were found using the Hamiltonian framework at spatial infinity which implements this antipodal map \cite{Troessaert:2017jcm,Henneaux:2018cst}. When the generalized BMS group can act, the boundary conditions at $\mathcal{I}^{+}_{\pm}$ are given by 
\begin{equation}
\begin{split}
C_{AB} &= (u+C_\pm) N_{AB}^{\text{vac}}-2D_A D_B C_\pm + q_{AB}D^E D_E C_\pm + o(u^0), \\
N_{AB} &= N_{AB}^{\text{vac}}+o(u^{-1}),
\end{split}
\end{equation}
as $u \rightarrow \pm \infty$. The quantity  $C_+(\theta,\phi) - C_-(\theta,\phi)$ is the supertranslation-invariant displacement memory field which sources the displacement memory effect \cite{Strominger:2014pwa}. These boundary conditions can be derived from the BMS orbit of Minkowski spacetime (i.e. the metric resulting from a general finite BMS diffeomorphism acted on Minkowski) whose Riemann-flat metric takes the exact form \cite{Penrose:1972xrn,Barnich:2016lyg,Compere:2016jwb}
\begin{equation}
\begin{split}
ds^2=-\frac{\mathring{R}}{2}du^2-2dr du+(r^2 q_{AB}+r C^{\text{vac}}_{AB}+\frac{1}{8}C^{\text{vac}}_{CD}C^{CD}_{\text{vac}}q_{AB})dx^A dx^B +D^B C_{AB}^\text{vac}dx^A du. 
\end{split}
\end{equation}
Here the vacuum shear is given by  $C^{\text{vac}}_{AB}=(u+C)N_{AB}^{\text{vac}}-2D_AD_B C+q_{AB}D^CD_C C$. The field $C(x^A)$ is either $C_\pm(x^A)$ depending on whether one considers the initial or final vacuum. This is the 4d analogue of the 3d metric \eqref{3dmetric}. Transitions between vacua are generated in general relativity by gravitational wave radiation. 

The leading and subleading soft graviton theorems are essentially the Ward identities of the BMS symmetries. One defines the total fluxes at $\mathcal I^+$ as 
\begin{equation}
F^+_\xi=Q_\xi \vert_{\mathcal I^+_+} -Q_\xi \vert_{\mathcal I^+_-} = \int^{+\infty}_{-\infty} du \partial_u Q_\xi . 
\end{equation}
The fluxes are similarly defined at $\mathcal I^-$. The quantization of the identities ``$F^+_\xi=$ the antipodal map of $F^-_\xi$'' are the leading and subleading soft graviton theorems, up to a switch from position basis to momentum basis \cite{Strominger:2017zoo}. Choosing the prescription $\bar N_A = N_A- u \partial_A \bar M+\frac{1}{4}C_{AB}D_C C^{BC}+\frac{3}{32}\partial_A(C_{BC} C^{BC})$ where $N_A$ is defined as in \cite{Barnich:2011mi}, the BMS flux algebra is given by \cite{Campiglia:2020qvc,Compere:2020lrt}
\begin{equation}    \label{algQ}
\begin{split}
&\{ F^+_{\xi(T_1)},F^+_{\xi(T_2) }\}=0, \\
&\{ F^+_{\xi(R_1)},F^+_{\xi(T_2)}\}=F^+_{\xi(\hat T)},\qquad \hat T=R_1^A \partial_A T_2-\frac{1}{2}D_A R^A_1 T_2 , \\
&\{F^+_{\xi(R_1)},F^+_{\xi(R_2)}\} = F^+_{\xi(\hat R)},\qquad \hat R^A=R_1^B \partial_B R^A_2 - R_2^B \partial_B R_1^A, 
\end{split}
\end{equation}
where $\{,\}$ is the standard Peierls bracket, which faithfully reproduces the BMS algebra of vector fields \eqref{alg}. Choosing the other prescription $\bar N_A= N_A- u \partial_A \bar M$ where $N_A$ is defined as in \cite{Barnich:2011mi} leads instead to the well-established geometrical definition of the angular momentum $Q_{\xi(R=\partial_\phi)}$ \cite{Thorne:1980ru,Ashtekar:1981bq,Landau:1982,Dray:1984rfa,Wald:1999wa,Barnich:2011mi}, see \cite{Compere:2021inq} for a discussion.

Other gauges for asymptotically flat spacetimes at null infinity lead to alternative residual gauge transformations, which can be associated to distinct charges. In harmonic gauge, several alternative charges have been proposed: it was shown that the subleading soft graviton theorems are associated with particular residual symmetries in harmonic gauge that crudely behave as $\xi^A \sim u R^A(\theta,\phi)+\dots $ \cite{Campiglia:2016efb}. It was also shown that the two infinite sets of canonical multipole moments \cite{Thorne:1980ru} are associated with two other subsets of residual gauge transformations in harmonic gauge, after a renormalization procedure and a prescription \cite{Compere:2017wrj}. Magnetically dual BMS supertranslations have also been defined \cite{Godazgar:2018qpq,Godazgar:2018dvh,Godazgar:2020kqd} which however might be trivial in the sense of associated with vanishing charges for standard asymptotically flat spacetimes \cite{Oliveri:2020xls}. At the time of finishing these lecture notes, a consistent large set of symmetries, $GL(\infty,\mathbb R)$, was found for asymptotically flat spacetimes at null infinity \cite{Guevara:2021abz,Strominger:2021lvk} based on the structure of the infinite set of soft theorems. However, it is not clear how to obtain this global symmetry group as asymptotic symmetry group. It seems that several coordinate systems and dualities will be required to exhaust
all possible charges of gravity in asymptotically flat spacetimes in order to find the largest possible asymptotic symmetry group.

\subsection{$4d$ scalar field - dual $U(1)$ Kac-Moody}

The role of asymptotic symmetries for scalar fields is an interesting story. Scalar fields, similarly to QED and gravity, admit soft theorems in which a scattering amplitude with an external massless scalar whose energy is taken to zero can be written as a soft factor times the scattering amplitude without the soft scalar. Correspondingly, they admit classically an infinite number of conserved charges whose Ward identities reproduce the soft theorems \cite{Campiglia:2018can}. While the standard formulation of scalar fields do not admit gauge freedom, and therefore asymptotic symmetries, dual formulations do admit asymptotic symmetries associated to local conserved charges \cite{Campiglia:2018see,Francia:2018jtb,Henneaux:2018mgn}. This case justifies the definition of the asymptotic symmetry group as the union of the asymptotic symmetries of all formulations of the same theory related by dualities.  

As a definite example, we consider the model \cite{Campiglia:2018see} of a  massless scalar field $\phi$ coupled to a massive scalar $\chi$ described be the Lagrangian density
\begin{eqnarray}
    {\cal L} &=& -\dfrac{1}{2}(\partial\phi)^2-\dfrac{1}{2}(\partial \chi)^2-\dfrac{1}{2}m^2\chi^2+\dfrac{g}{2}\phi\chi^2.
\end{eqnarray}
The equations of motion following from this Lagrangian are
\begin{eqnarray}\label{eq:interacting-scalar-eom}
    \Box \phi = -\dfrac{g}{2}\chi^2,\qquad \Box \chi -m^2 \chi=-g \phi \chi . 
\end{eqnarray}
The metric is the Minkowski metric in retarded spherical coordinates $ds^2=-du^2-2du dr+r^2 q_{AB}dx^A dx^B$. The massive scalar reaches the future timelike boundary $i^+$ modelled as $EAdS^+_3$ while the massless scalar reaches the null boundary $\mathcal I^+$. A combined analysis of both boundaries is therefore required. At $\mathcal I^+$, the asymptotic fall-off conditions are given by 
\begin{eqnarray}
    \phi(u,r,x^A) =\frac{\varphi(u,x^A)}{r}+O(r^{-2}),\quad \varphi_\pm(x^A) := \lim_{u\to \pm\infty}\varphi(u,x^A). 
\end{eqnarray}
The boundary $EAdS^+_3$ is obtained as $\tau\rightarrow\infty$ in the hyperbolic foliation of Minkowski $ds^2=-d\tau^2+\tau^2 \left(  \frac{d\rho^2}{1+\rho^2}+\rho^2 q_{AB}dx^A dx^B \right)$. The fall-off conditions are given by 
    \begin{align}
        -\frac{g}{2}\chi^2 & = \frac{j(\rho,x^A)}{\tau^3}+ O(\tau^{-4}), \\
        \phi &= \frac{\phi_{i^+}(\rho, x^A)}{\tau}+O(\tau^{-2}). 
    \end{align}
The field equations \eqref{eq:interacting-scalar-eom} imply at leading order in the limit $\tau \rightarrow \infty$ that $\Box (\frac{\phi_{i^+}(\rho, x^A)}{\tau})= \frac{j(\rho,x^A)}{\tau^3}$. The junction condition between $EAdS^+_3$ and $\mathcal I^+$ can be written as 
\begin{equation}
\phi(\rho \rightarrow \infty,x^A)=\varphi_+(x^A) .     
\end{equation}
The value $\varphi_+(x^A) $ can then be computed from a 3d-2d bulk-to-boundary propagator sourced by the boundary current $j(\rho,x^A)$ \cite{Campiglia:2018see}: $\varphi_+(x^A)=\frac{1}{4\pi}\int_{i^+}d^3 Y \frac{j(Y)}{Y \cdot x}$. Here $x^\mu = (1,x^A)$, $Y^\mu = (\sqrt{1+\rho^2},\rho y^A)$ is a representation of points on $i^+$, defined so that $Y^2=1$, with $d^3Y$ the measure on $i^+$.

In terms of these quantities the Noether charge one may guess from the scalar soft theorem is \cite{Campiglia:2018can}
\begin{eqnarray}\label{eq:guessed-scalar-noether-charge}
    Q_\lambda^+&=&4\pi\int_{\cal I^+_-}\lambda(x^A)\varphi_-(x^A)\\
    &=& \underbrace{-4\pi\int_{\cal I^+} \lambda(x^A)\partial_u\varphi(u,x^A)}_{\text{Soft Part}} + \underbrace{\int_{S^2} d^2S \lambda(x^A)\int_{i^+} d^3Y\frac{j(y)}{Y(y)\cdot x}}_{\text{Hard Part}}. 
\end{eqnarray}
As in the electromagnetic and gravitational case, the charge has both a soft and a hard part.

A completely analogous construction can be carried out on ${\cal I}^-\cup i^-$ and compatibility between the two boundaries, ensuring charge conservation, follows from an antipodal identification of the scalar field near $i^0$, namely $\phi|_{\cal I^+_-}(\hat{x})=\phi|_{\cal I^-_+}(-\hat{x})$ \cite{Campiglia:2018can} where $\hat x$ is the unit normal to the celestial sphere.

Let us now discuss how one can more fundamentally define this charge by promoting it as a Noether charge associated with asymptotic symmetries after employing a duality \cite{Campiglia:2018see}. On the one hand, a massless scalar such as $\phi$ is dual to a $2$-form gauge field ${\cal B}$ with field strength ${\cal H}=d{\cal B}$ related to $\phi$ as ${\cal H}=\star d\phi$, in terms of which the massless scalar free action is written as
\begin{eqnarray}
    S_{\cal B} = -\frac{1}{2}d\phi\wedge\star d\phi = -\dfrac{1}{2}{\cal H}\wedge \star{\cal H},\quad {\cal H}:=d{\cal B}.
\end{eqnarray}
The massive field $\chi$ on the other hand is replaced by a collection of massive point particle worldlines interacting with ${\cal H}$
\begin{eqnarray}
    S_{\rm pp} + S_{\rm int} = -m\sum_i \int d\tau_i +\dfrac{g}{12m}\sum_i d\tau_i \int^{\tau_i}d\tau'\dfrac{dx^\mu}{d\tau'}\varepsilon_{\mu\nu\rho\sigma}{\cal H}^{\nu\rho\sigma}(x_i(\tau')).
\end{eqnarray}
The resulting action is therefore $S = S_{\cal B}+S_{\rm pp}+S_{\rm int}$ and since it depends just on ${\cal H}$ it has a gauge symmetry $\delta_\alpha{\cal B}=\alpha$ with any closed form $d\alpha=0$. Using $\mathcal H = d\mathcal B=\star d\phi$ one directly obtains $\alpha\wedge \star \mathcal H=d(\phi \ \alpha)$. 

The Noether charge following from this symmetry is evaluated from the covariant phase space algorithm to be
\begin{eqnarray}
    Q_\alpha^+ = \int_{\cal I^+_-}\phi\, \alpha = \underbrace{\int_{\cal I^+}\alpha\wedge \star {\cal H}}_{\text{Soft Part}}+\underbrace{\int_{i^+}\alpha\wedge \star {\cal H}}_{\text{Hard Part}} . 
\end{eqnarray}
If the $2$-form $\alpha$ is not proportional to the $S^2$ volume form the charge vanishes and the gauge transformation is trivial. The non-trivial gauge transformations are of the form $\alpha = \lambda(x^A)\epsilon$ where $\epsilon$ is the $S^2$ volume form and therefore the covariant phase space formalism applied in this manner really recovers the charge (\ref{eq:guessed-scalar-noether-charge}) guessed from the scalar soft theorem. Since all charges mutually commute under the Peierls bracket (because $\delta_\lambda \mathcal H = 0$), the algebra is abelian: it is a $U(1)$ Kac-Moody algebra, $\{Q_\lambda, Q_{\lambda'}\}=0$. 

\section{Frontiers in gravity}

\subsection{Higher dimensions}

Asymptotically flat spacetimes in dimensions higher than 4 are not obvious generalizations of the cases $d=3,4$. First, there is a qualitative distinction between even and odd spacetime dimensions which can be simply explained as follows. Massless fields reaching future null infinity are governed by the retarded Green function which is a solution to 
\begin{equation}
    \partial_t^2 G - \sum_i^d \partial_i^2 G = \delta (t-t_0)\delta (\Vec{r}-\vec{r_0}).
\end{equation}
In even dimensions, the retarded Green function takes the form $G(t,r)= \frac{1}{4\pi} \big(-\frac{1}{2\pi r} \partial_r \big)^{\frac{d-3}{2}}\big(\frac{\partial(t-r)}{r}\big)$ which can be expanded for large radius as a polynomial expansion in $r$. Instead, in odd dimensions, the retarded Green function takes the form $G(t,r)=\frac{\theta(t)}{2\pi}\big(-\frac{1}{2\pi r}\partial_r \big)^{\frac{d}{2}-1}\big(\frac{\theta(t-r)}{\sqrt{t^2-r^2}}\big)$ which admits non-analytic half-integer fall-off in $r$ in the large $r$ limit \cite{Dai:2013cwa}. Penrose's conformal methods that are based on analyticity at null infinity are therefore inapplicable in odd spacetime dimensions. Asymptotic analyses can still be performed using the gauge-fixing approach where explicit coordinates allow to treat even non-analytic asymptotic behavior. 

The asymptotic symmetry group of asymptotically flat spacetimes at null infinity in even $d>4$ has been first obtained to be the Poincar\'e group and nothing else using conformal compactification techniques   \cite{Hollands:2003ie,Tanabe_2009,Tanabe_2010,Tanabe_2011}. However, a closer analysis revealed further structure. 

A noticeable feature of higher dimensional gravity is that the Newtonian 
potential falls off as $r^{3-d}$ at null infinity in the limit $r\rightarrow \infty$ while the radiation falls off much slower as $r^{1-\frac{d}{2}}$. Only in $d=4$ these two fall-off conditions agree but in higher dimensions, the radiation is leading. The asymptotic symmetries are not associated with the radiative fall-off, but instead with the Newtonian fall-off. 

The standard translations are given by $\xi = T(x^A)\partial_u+\dots $ where $T(x^A)=1 $ for time translations and $T$ obeys $(\Delta +d-2)T(x^A)=0$ for spatial translations where $\Delta=D_A D^A$ is the Laplacian. The charges associated with these asymptotic symmetries are  the finite momenta. The Bondi mass aspect can be defined in Bondi gauge as the following component of the Weyl tensor at $\mathcal I^+$: $m=r C_{ruru}$. The supermomentum charges or supermomenta are then defined as $Q_\xi = \int_S d\Omega \,  m(u,x^A)T(x^A)$ where $T(x^A)$ is an arbitary function over the sphere. Quite naturally, one can promote these charges as Noether charge associated with supertranslations of the form $\xi = T(x^A)\partial_u+\dots $ \cite{Awada:1985by,Barnich:2006av,Kapec:2015vwa}.

Another development arises from the displacement memory effect \cite{Mao:2017wvx,Conde:2016rom,Pate:2017fgt}. The leading displacement memory also appears at the same order $r^{3-d}$ as the Newtonian order which is again lower than radiative orders $r^{1-\frac{d}{2}}$ for $d>4$ contrary to $d=4$. One can identify a component of the metric in the radial expansion close to $\mathcal I^+$ in harmonic gauge lets say $\varphi$ such that the leading displacement memory is sourced by the difference between the future and past values of the field at null infinity: $M= \varphi\vert_{\mathcal I^+_+}-\varphi\vert_{\mathcal I^+_-}$. By definition of memory, i.e. a transition between two vacua, there exists a gauge transformation such that $\delta_\xi \varphi\vert_{\mathcal I^+_+}=\delta_\xi \varphi\vert_{\mathcal I^+_-}$ since in a given asymptotic vacuum, the field $\varphi$ is pure gauge while the source of the memory $M$ is gauge invariant. After analysis, this gauge transformation reads in harmonic gauge as $\xi = r^{4-d}\Delta(\Delta+d-2)T(x^A)\partial_u+\dots$. It is in a sense a subleading supertranslation. Its associated charge is vanishing as $Q_\xi=O(r^{4-d})$ and subleading supertranslations are therefore not asymptotic symmetries in the standard sense of being associated with finite charges. 

The status of super-Lorentz transformations in higher dimensions is murkier. The natural candidate as super-Lorentz group is $\text{Diff}(S^{d-2})$. The angular momentum in the Myers-Perry black hole solution (the generalization to higher dimensions of the Kerr black hole) appears at order $r^{2-d}$ \cite{Myers:1986un}. After renormalization, we expect that one can define finite surface charges associated with generators $\xi=R^A(x^B)\partial_A +\dots$ of the form $Q_\xi=\int d\Omega N_A(u,x^B)R^A(x^B)$ \cite{Capone:2019aiy,Colferai:2020rte,Campoleoni:2020ejn,Fiorucci:2020xto,Capone2021}. Subleading memory effects have not yet been studied. 

Asymptotic symmetries and charges of scalars theories, Maxwell theory, Einstein gravity and linear higher spin fields were further studied in both even and odd dimensions in \cite{Campiglia:2017xkp,Campoleoni:2017qot,He:2019jjk,Esmaeili:2019hom,Henneaux:2019yqq,Colferai:2020rte,Campoleoni:2020ejn}.

\subsection{Supersymmetric extensions}

Given the occurrence of infinite dimensional symmetry groups as bosonic asymptotic symmetry groups, it is very natural to ask what the supersymmetric extensions of these groups are and which of these extensions are realized as asymptotic symmetry groups of supersymmetric gauge and gravity theories. The motivation comes essentially from the fact that all known exact holographic dualities involve a high level of supersymmetry. Finding supersymmetric extensions of the BMS group brings celestial holography closer to achieving its goals. We will restrict our comments only on the cases of gravity in 3 and 4 dimensions. 

\subsubsection{$\mathcal N=1$ super-BMS$_3$ from $\mathcal N=1$ 3d supergravity}

Extended $\mathcal N>1$ BMS$_3$ algebras have been formulated and have been realized within extended supergravities at future (or past) null infinity. The $\mathcal N=1$ super-BMS$_3$ algebra realized in terms of surface charges under the Peierls bracket is the union of its bosonic part \cite{Barnich:2006av} which we already discussed in Eq. \eqref{BMS3}, 
\begin{equation}
\begin{split}
     i\{\mathcal{P}_m, \mathcal{P}_n\} &= 0; \\ i\{\mathcal{J}_m ,\mathcal{J}_n\}& = (m-n)\mathcal{J}_{m+n} + \frac{c_1}{12}m^3\delta_{m+n,0}, \qquad\qquad  c_1 = 0; \\ i\{\mathcal{J}_m, \mathcal{P}_n\} &=  (m-n)\mathcal{P}_{m+n}+ \frac{c_2}{12}m^3\delta_{m+n,0}, \qquad \qquad  c_2 = \frac{3}{G},
\end{split}
\end{equation}
and its fermionic part ($Q_n$ is Grassmann odd) \cite{Bagchi:2009ke,Mandal:2010gx,Barnich:2014cwa}
\begin{equation}
\begin{split}
     i\{\mathcal{P}_m, \mathcal{Q}_n\} &= 0, \\ i\{\mathcal{J}_m ,\mathcal{Q}_n\} &= (\frac{m}{2}-n)\mathcal{Q}_{m+n}, \\
     \{\mathcal{Q}_m, \mathcal{Q}_n\} &=  \mathcal{P}_{m+n}+ \frac{c_2}{6}m^2\delta_{m+n,0}, \qquad  c_2 = \frac{3}{G}.
\end{split}
\end{equation}
It is realized as asymptotic symmetry group of the simplest theory of 3d supergravity \cite{Barnich:2014cwa}. Larger supersymmetric extensions have been considered in \cite{Banerjee:2016nio,Lodato:2016alv,Banerjee:2017gzj,Fuentealba:2017fck,Poojary:2017xgn,Banerjee:2018hbl,Valcarcel:2018kwd,Caroca:2019dds,Banerjee:2019lrv,Banerjee:2019epe,Chernyavsky:2019hyp}.

\subsubsection{$\mathcal N=1$ super-BMS$_4$ from $\mathcal N=1$ 4d supergravity}

We discussed the BMS$_4$ asymptotic symmetry algebra of asymptotically flat spacetimes at future (or past) null infinity in Section \ref{BMS4}. We now extend the smooth supertranslations and the $\text{Diff}(S^2)$ super-Lorentz generators to arbitrary functions of the stereographic coordinates $z,\bar z$ which include poles on the sphere. We then consider the subalgabra where the super-Lorentz generators are restricted to meromorphic and anti-meromorphic functions such that $T=T(z,\bar z)$ but $R^z=R^z(z)$ and $R^{\bar z}=R^{\bar z}(\bar z)$. The asymptotic symmetry vector fields depending upon these generators can be extended in modes as
\begin{equation}
    P_{k,l} = z^{k+\frac{1}{2}}\bar{z}^{l+\frac{1}{2}} \partial_u +\dots, \qquad     l_m = -z^{m+1}\partial_z+\dots , \qquad \bar{l}_m = -\bar{z}^{m+1}\partial_{\bar z}+\dots,  
\end{equation}
where the dots are tuned such that the vectors preserve Bondi gauge and $k,l,m \in \mathbb Z$. They obey the so-called extended BMS$_4$ algebra \cite{Barnich:2009se}
\begin{equation}
    \begin{split}
        [l_m, l_n] &= (m-n)l_{m+n}, \qquad  [\bar{l}_m, \bar{l}_n] = (m-n)\bar{l}_{m+n}, \\
        [l_m, P_{k,l}] &= (\frac{1}{2}m-k)P_{m+k,l}, \qquad  [\bar{l}_m, P_{k,l}] = (\frac{1}{2}m-l)P_{k,m+l}, \\
        [l_m, \bar{l}_n] &=0, \qquad  [P_{k,l}, P_{o,p}] = 0. 
    \end{split}
\end{equation}
One can supplement this bosonic algebra with the following fermionic part in terms of the Grassman odd generators $G_m$, $\bar G_{m}$, $m \in \mathbb Z$ to obtain the graded Lie algebra known as the $\mathcal N=1$ super-BMS$_4$ algebra
\begin{equation}
    \begin{split}
        \{G_m, \bar{G}_n\} = P_{m,n}, \hspace{10pt} \{G_m, G_n\} = \{\bar{G}_m, \bar{G}_n\} = 0 \hspace{10pt} [L_m, \bar{G}_n] = [\bar{L}_m, G_n] = 0 ,\\
        [P_{k,l}, G_n] = [P_{k,l}, \bar{G}_m] = 0, \hspace{10pt} [L_m, G_k] = (\frac{1}{2}m-k)G_{m+k}, \hspace{10pt} [\bar{L}_m, \bar{G}_l] = (\frac{1}{2}m-l)\bar{G}_{m+l}.
    \end{split}
\end{equation}

Such algebra is realized as asymptotic symmetry algebra of $\mathcal N=1$ supergravity at null infinity \cite{Awada:1985by,Lysov:2015jrs,Avery:2015gxa,Fotopoulos:2020bqj}. Its subalgebra without super-Lorentz generators but Lorentz generators ($m,n=-1,0,1$; $k,l \in \mathbb Z$) can also be realized at spatial infinity \cite{Fuentealba:2021xhn}. Note that another inequivalent extension of the BMS$_4$ algebra with only four fermionic generators also exists \cite{Awada:1985by,Henneaux:2020ekh}. 

\vspace{10pt }
 \centerline{\bf Acknowledgements}

G. C. would like to acknowledge the organizer of the PhD school on Celestial Holography at Princeton, 2021 where this lecture took place. G. C. is Senior Research Associate of the F.R.S.-FNRS. and acknowledges support from the FNRS research credit No. J003620F and the IISN convention No. 4.4503.15. L. P. de Gioia would like to acknowledge the organizer of the school on Celestial Holography and also acknowledges support from Conselho Nacional de Desenvolvimento Cient\'{ı}fico e Tecnol\'{o}gico (CNPq, process number 140725/2019-9). P. B. A would like to acknowledge the organizer of the school on Celestial Holography and Alok Laddha and Anupam A. H. for numerous valuable discussions on all BMS things. P. B. A is supported in part by a grant to CMI from the Infosys foundation.


\providecommand{\href}[2]{#2}\begingroup\raggedright\endgroup

\end{document}